\documentclass[a4paper,12pt]{article}
\usepackage{pratex}
\usepackage{epsfig}
\usepackage{rotating}
\date{} 
\pranum{00/04} 
\begin{document}
\begin{titlepage}
\title{Can We Observe the Quark Gluon Plasma in Cosmic Ray Showers ?}
\author{Jan Ridky\footnote{e-mail:ridky@fzu.cz}}
\address{Institute of Physics, Acad. of Sci. of the Czech Rep.} 

\begin{abstract} 
The possibility of detection of some features of high energy particle
interactions with detectors placed at medium depths underground through 
studies on high energy muons is studied. These muons carry information 
about the early interactions occurring during the development of the 
hadron cascade near the top of the atmosphere. They might reveal the
effects resulting from creation of quark gluon plasma in interactions 
of ultra high energy cosmic ray iron nuclei with air nuclei. 
\end{abstract}
\note{submitted to {Astroparticle Physics}} 
\end{titlepage}

\section{Introduction} 
\label{secI}

The cosmic ray composition near the knee region ($5 \times 10^{14}~eV$
up to almost $10^{17}~eV$) is a challenging problem \cite{WATSON}.
Recent measurements by KASCADE \cite{KASKADE1}
have suggested an increased proportion of heavier nuclei in cosmic ray flux
at these energies. However, these results are still far from conclusive 
as the results derived from electron, muon and hadron data differ and 
the composition measurements from Cerenkov arrays show even decreasing 
trend \cite{DICE},\cite{CASA-BLANCA}.
Although more precise determination of the chemical composition is not 
yet feasible, the presence of $Fe^{56}$ nuclei at $\sim 10-20\%$ level 
in the primary flux at $\sim 10^{15}~eV$ seems quite plausible. 
This result presents an interesting opportunity to observe
iron~nitrogen and iron~oxygen nucleus-nucleus interactions at energies
beyond the possibilities of terrestrial accelerators. Though
both $N^{14}$ and $O^{16}$ are light nuclei\footnote{In principle
interactions with $Ar^{40}$ may also happen but these are suppressed by
a factor larger than $\sim$ 100 according to CORSIKA simulations.}, 
the number of participating nucleons in head-on collisions with iron 
nuclei may go upto several tens. Thus, given the very high energy 
per nucleon available with the incoming iron nuclei, the formation of
quark-gluon plasma seems to be a likely possibility in a fraction of
the primary interactions. 
Although the baryon density is only half of the density in Pb-Pb interactions 
(due to the interaction with lighter nuclei), the energy density $\epsilon$ 
can reach sufficiently high value. This will be shown in section~\ref{secS}
The conditions in which such interactions are taking place are unique.
However they provide an opportunity to investigate some characteristics of 
quark-gluon plasma (QGP) which are not available presently in
terrestrial experiments. The interactions of iron nuclei take place at
altitudes higher than 20 km and secondary charged pions have large 
probability to decay into muons due to the thin air at these altitudes. 
While the decay products of neutral pions undergo further interactions 
leading to the development and absorption of the electron-photon casacde 
in the middle and lower atmosphere, the muons, in some sense, decouple 
from the shower development and are subjected only to energy loss 
through ionization and  multiple scattering. Moreover, the large Lorentz 
factor helps in keeping the secondary particles collimated within a
narrow cone. As a result, there is a possibility to observe differences 
in the characteristics of the muon component in some extensive air
showers from what one would expect on the basis of standard models of
high energy interactions. 

The fact that the hard component of muons in extensive air showers
carries information on the pion production in the initial stages of the
shower development is of great importance. Hence, aside from QGP, it is
also interesting to study the sensitivity of energetic muons to the
details of the primary interaction dynamics. In this regard the QGP
effects observed in the muon component of air showers will also indicate 
the extent of the above mentioned sensitivity. Of course it is necessary
to determine in more quantitative way the notion of 'hard component' 
in terms of energy and the methodology to observe it. One simple method is 
to place a large area high resolution muon detectors underground at 
such depth that the overburden will filter out the less energetic muons 
and preferentially provide the information on the initial interaction.

Due to the absence of a comprehensive theory on the formation and 
evolution of QGP and also due to the lack of solid experimental evidence 
there are many phenomenological models with quite different and sometimes 
even opposite predictions. The same is also true for results on secondary
particles multiplicity. There are predictions of a decrease of
multiplicity \cite{ESKOLA} in high energy nuclear collisions. On the other
hand, many statistical thermal models predict increase of multiplicity 
\cite{ARMESTO}. Anyway the secondary particle multiplicity is an
important variable and a change in the interaction dynamics may be
observable through change in the characteristics of particle production, 
depending on the size of the effect and the sensitivity of measurements.
The aim of this paper is to study such phenomena through detailed 
simulations and to learn about conditions necessary to observe at 
least some of the features which would reveal the formation of QGP 
in cosmic ray interactions in the atmosphere. To achieve this, a very 
simple model of QGP formation is presented in section~\ref{secS} This
model is then used to simulate the formation of QGP and the results are 
compared with the outcome of programs currently used for simulations of 
extensive air showers. The analysis of simulated data and comparisons 
are presented in section~\ref{secR} The conclusions are discussed in 
the final section.

\section{Quark Gluon Plasma Simulation} %
\label{secS}

The standard package used for simulation of cosmic ray interactions in the
atmosphere is the program CORSIKA \cite{CORSIKA} supplemented by
several models of high energy interactions of elementary particles. The
properties of these models have been extensively studied in
\cite{KASKADE2},\cite{KASKADE3}. Good agreement with experimental 
data of KASCADE has been achieved using QGSJET \cite{QGSJET} and VENUS
\cite{VENUS} models with QGSJET giving somewhat better results. VENUS 
predicts slightly larger number of muons per shower than QGSJET at 
energies in the range, $10^{14}~eV - 10^{16}~eV$. As this feature might 
blur or hide the muon component contributed by the QGP, the VENUS model 
has been used for comparisons while searching for the observable effects
due to the QGP. Another reason for this choice is the availability of 
good documentation for VENUS and the fact that QGP simulations 
supplemented by VENUS simulations were possible without any intervention 
with the VENUS code. 

\begin{figure}\centering
\epsfig{file=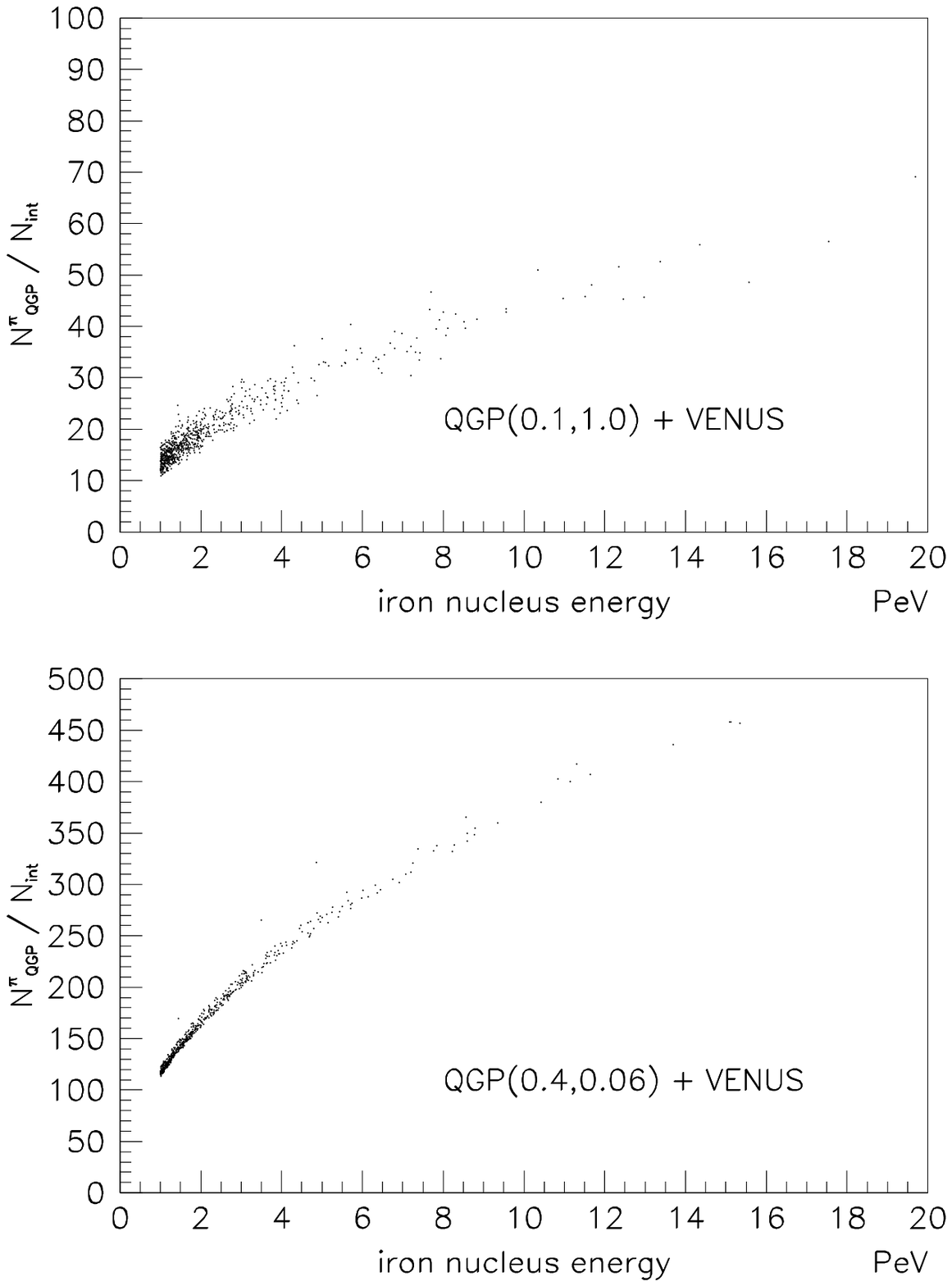,width=15cm,height=20cm}
\caption{Number of pions (both $\pi^{\pm}$ and $\pi^{0}$) produced per
  NN interaction.} 
\label{fig:npipernn}
\end{figure}

The key factor in QGP modelling is the number of nucleons, $N_{int}$,
participating in the first interaction of an iron nucleus with the 
atmospheric nucleus. If $N_{int}$ is larger than a certain limit, 
$N_{hot}$, nucleons are supposed to melt down into QGP. The values for 
the fraction of melted nucleons used in simulations were 10 \%, 20 \% 
and 40 \% and the corresponding threshold values of $N_{int}$ were chosen 
to be 13, 15 and 19 nucleons. We shall denote this fraction as
${\cal R}_{h}=N_{hot}/N_{int}$. Thus, in the case of ${\cal R}_{h}= 0.1$,
only two nucleons can melt. For values of $0.2$ and $0.4$, this minimum
is 3 and 8 nucleons respectively. The aim was to keep the minimum $N_{int}$ 
leading to QGP production to be above 10 nucleons. In the process of 
simulation, $N_{hot}$ nucleons are taken aside and rest of the particles 
participates further in collision simulated with the VENUS code. The total 
energy of $N_{hot}$ nucleons, denoted $E_{hot}$, is used to produce again 
$N_{hot}$ nucleons, but this time with the kinetic energy generated according 
to the Boltzmann distribution and isotropically in the overall centre of 
mass system. The remaining $E_{hot}$ is used for production of $\pi^{\pm}$ 
and $\pi^{0}$, again with Boltzmann distributed kinetic energy and 
isotropically in the centre of mass system. The pion production is stopped 
when the $E_{hot}$ is reduced to $2.5~GeV$. Afterwards this remnant cluster 
is passed to CORSIKA for further processing. For each of the three values 
of ${\cal R}_{h}$ considered, the pions were generated with Boltzmann 
distribution parameters corresponding to the mean pion momenta 
$<p^{\pi}\!\!>$ values of $60~MeV$, $120~MeV$, $300~MeV$ and $1~GeV$. 
The wide range of the values of the fraction, ${\cal R}_{h}$, together with 
a range of possible values of the mean momentum of generated pions are 
intended to account for uncertainties in the description of the QGP freeze 
out. The properties of eventual QGP formed in light nuclei interactions 
are also unknown. For example, the temperature might not be high enough to 
produce $c\bar{c}$ or even $s\bar{s}$ quark pairs. 
Also the isotropy of pions radiated during the
freeze out is not assured and the mean momentum along the shower
axis may differ from momenta in the transverse plane. However, it
turns out that the results are not sensitive to this. The results of 
simulations with $<p^{\pi}_{L}>=300~MeV$ and $<p^{\pi}_{T}>=30~MeV$ do
not differ significantly from those with overall $<p^{\pi}>=300~MeV$.
Hence, for the sake of simplicity, the pion production is considered
to be isotropic. The mean number of all kinds of pions produced under 
different conditions in this simple approach are
given in the Table~\ref{tab:hotpions}. The number of pions produced
per nucleon-nucleon interaction and its dependence on the primary energy
are shown in  Fig.~\ref{fig:npipernn}. The two extreme cases - one
with maximum pion production, i.e. 40 \% of $N_{int}$ nucleons melting
down and pions evaporating with mean momentum of $60 MeV$ (henceforth we
shall abbreviate this as $QGP(0.4,0.06)$), and the other with minimum
pion production ($QGP(0.1,1.0)$) are shown.
\begin{table}\centering
\begin{tabular}{|c|c|c|c|}   \hline
$<p^{\pi}\!\!>$ & \multicolumn{3}{c|}{${\cal R}_{h} \;\; ratio$ } \\ \cline{2-4}
$MeV$      &     0.1     &     0.2      &     0.4      \\ \hline
60         &  1492 (880) &  2743 (1717) &  6764 (3802) \\ \hline
120        &  1389 (791) &  2610 (1634) &  6332 (3568) \\ \hline
300        &  1074 (680) &  2165 (1398) &  4918 (2614) \\ \hline
1000       &  ~674 (393) &  1260 ~(825) &  3093 (1808) \\ \hline
\end{tabular}
\caption{Mean numbers of pions (both $\pi^{\pm}$ and $\pi^{0}$) produced 
in QGP. Figures given in the parentheses represent the R.M.S. values.} 
\label{tab:hotpions}
\end{table}
All the data sets under consideration here were generated from 2000 iron 
nucleus initiated showers with the primary energy within the range 
$10^{15}~eV - 2.~10^{17}~eV$ (the upper limit is the one recommended
as a limit of reliability of the VENUS program \cite{CORSIKA}). The
spectral index of the energy distribution was $-2.7$ \cite{KASKADE1}.
The primary zenith angle was fixed to be $0$ degrees while the azimuth
angle and the altitude of the initial interaction were left free.
The mean primary energy of $2.2~10^{15}~eV$ is relevant to all results 
presented in the paper if not stated otherwise. 

The chosen primary energy interval should not be taken as a prediction 
for QGP formation at any specific energy. The primary energy range was 
determined essentially by the range of validity of the VENUS model and by 
flux considerations, i.e., the experiment with realistic area should be able
to observe the effects. This point will be further discussed in 
section~\ref{secD} The kinematical conditions and the dynamics of parton 
interactions as described by VENUS allow the QGP formation at these energies. 
One can estimate the energy density $\epsilon$ using the Bjorken's 
expression \cite{BJORKEN}

\[ \epsilon = {{\Delta E / \Delta y}\over{\pi R^{2} \tau}}, \]
where $R$ is the radius of the smaller nucleus, $\tau$ is the
formation time ($1 \div 2~fm/c$) and $E$ is the c.m.s. energy of
final state particles in the rapidity unit interval near $y_{cms} = 0$. 
With the transition temperature in a two-flavour QGP in the range, 
$150~MeV < T_{c} < 200~MeV$, the corresponding critical energy densities 
are $0.8~GeVfm^{-3} < \epsilon_{c} < 2.5~GeVfm^{-3}$. The VENUS
simulations give for $Fe-N$ interactions at $E_{inc} = 10^{15}~eV$
$\epsilon = 3.3~GeVfm^{-3}$ with impact parameter $b < 2~fm$
and $\tau = 1~fm/c$ ($\epsilon = 3.5~GeVfm^{-3}$ for $b~<~1~fm$). At energy
$E_{inc} = 10^{16}~eV$, the energy densities reach the values 
$\epsilon~=~4.9~GeVfm^{-3}$ for $b < 2~fm$ and $\epsilon = 5.4~GeVfm^{-3}$ 
for $b < 1~fm$. Thus even with some uncertainty in the value of $\tau$, the  
creation of QGP in the iron-air interactions at the ``knee'' energies is 
not excluded, at least according to VENUS. The lower part of the primary 
energy interval is already accessible at RHIC. However, the collider 
experiments explore mainly the central rapidity region while the cosmic 
ray experiments are sensitive to particle production in the very forward 
region. Thus the two approaches are in a sense complementary.

\section{Quark Gluon Plasma Effects} %
\label{secR}

The main detectable quantities used to describe extensive air showers
are the electromagnetic and muon sizes and corresponding lateral
distribution functions (LDF). These quantities are related as follows,
e.g. for the muon size $N_{\mu}$,

\[ N_{\mu} = \int^{r_{2}}_{r_{1}} 2\pi \rho_{\mu}(r) dr, \]
where $\rho_{\mu}(r)$ is the density of particles in 
the detection plane and $r$ is the distance from the shower axis.

\begin{figure}\centering
\epsfig{file=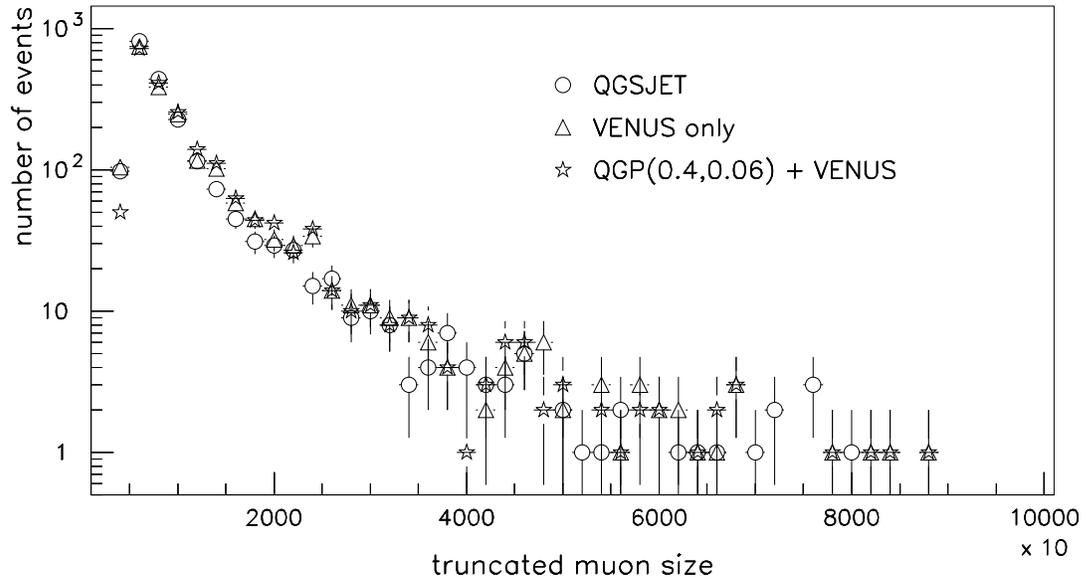,width=15cm,height=9cm}
\caption{Muon size truncated between 30 and 230 meters from the shower
  centre at the ground level.} 
\label{fig:musize}
\end{figure}

About one third of pions originating
from QGP are $\pi^{0}$. Their decay products undergo further
interactions in the atmosphere and they will disperse among other
particles of electromagnetic nature. As the electromagnetic
size is typically much larger than the muon size, the QGP effects
should be easier to detect in the muon component of extensive air
showers. However, the
muon sizes of showers at the ground level do not reveal any differences.
The truncated muon sizes $N^{tr}_{\mu}$ are compared in Fig.~\ref{fig:musize}.
The truncation corresponds more to the reality of the experimental conditions 
where the measurements near the shower core are hampered by systematics
caused by punch-through and depend on the details of detector design. 
The kinetic energy cutoff used here for pions is $E_{kin} > 300 \: MeV$.
The results obtained from simulations with QGSJET as well as VENUS but 
without QGP formation are compared in the figure with simulations of QGP 
supplemented by VENUS in the case of the most abundant pion production, 
$QGP(0.4,0.06)$.

\begin{figure}\centering
\epsfig{file=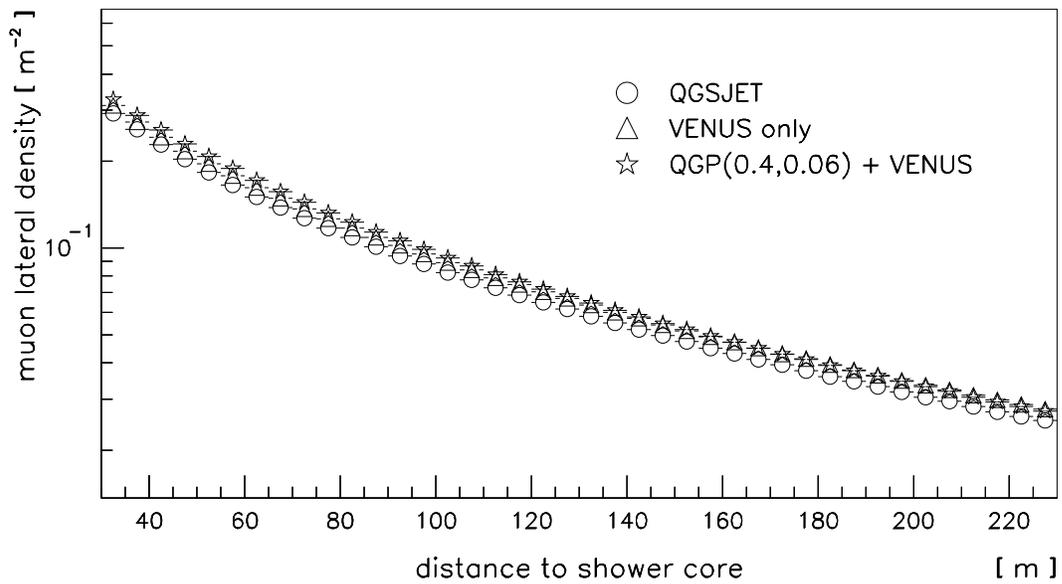,width=15cm,height=9cm}
\caption{Lateral distribution of muon densities on the ground.} 
\label{fig:gldf}
\end{figure}

Even the more detailed lateral distribution functions shown in 
Fig.~\ref{fig:gldf}
are almost identical and do not reveal any features which would allow us to 
observe effects corresponding to QGP formation. The events with QGP have 
slightly larger LDF values in the region $50 \div 100$ meters than the
VENUS simulation. However, this difference is not measurable and
the differences between QGSJET and VENUS are larger than
any QGP effects. The differences for events with other values of 
${\cal R}_{h}$ and $<p^{\pi}\!\!>$ are similar or even smaller.

As a consequence of kinematics, the muons originating from QGP formed
in the initial interaction of iron nucleus with air have mostly
higher energies than the muons produced lower down in the atmosphere. 
Thus detection of hard muons in cores of extensive air showers should 
reveal differences between cases with and without QGP production. 
To estimate this effect, the data are further analysed with 
detectors at a medium depth underground. Our somewhat simplified 
virtual experiment is placed at a depth of 110 m with 100 m thick 
overburden made of standard rock 
($A = 22, Z = 11$ and $\rho = 2.65 g/cm^{3}$). The
underground cavern with the 10 m high ceiling is supposed to have
detection area of $100 \times 100 \; m^{2}$. The tracking of particles 
generated by CORSIKA through the rock was done using GEANT \cite{GEANT}. 
The rock filters out the muons with momenta smaller than $53~GeV$.

\begin{figure}\centering
\epsfig{file=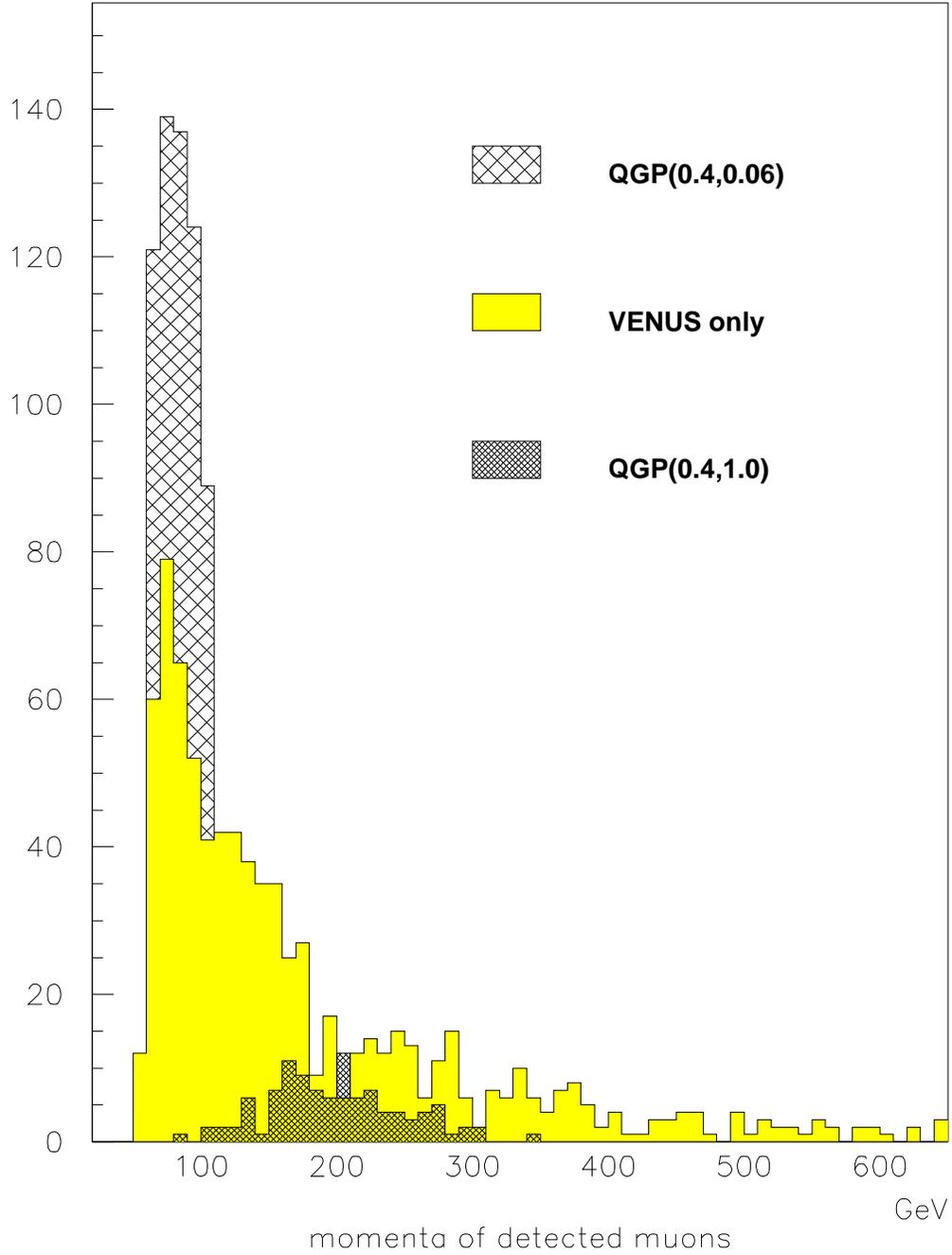,width=15cm,height=20cm}
\caption{Momentum distributions of muons from QGP and from standard 
production. Altitude of interaction is $33.1~km$, $E_{inc}=2.25~10^{15} eV$, 
$N_{int}=36$ and $N_{hot}=14$.} 
\label{fig:qgmevtmu}
\end{figure}

The muons penetrating underground have an average momentum around 
$280~GeV$ but the distribution has a very long tail. To illustrate the 
contribution of QGP originated muons relative to standard muon production, 
Fig.~\ref{fig:qgmevtmu} 
shows the momentum distribution of muons produced in a typical event
with $E_{inc}=2.25~10^{15} eV$, which is close to the average primary
energy, and with the height of the primary interaction at $33.1$ kilometres, 
corresponding to the maximum of the distribution. In this particular event
$N_{int}=36$, which is higher than the average. The reason for this 
selection was to obtain at ${\cal R}_{h}=0.4$ high QGP muon multiplicity. 
The superimposed contributions from $QGP(0.4,0.06)$ and 
$QGP(0.4,1.0)$ show strong dependence of QGP muon momenta on the
initial conditions of pion production. The underground detection
area is sensitive only to a small fraction of muons coming from QGP pion
decays as can be seen from Fig.~\ref{fig:qgmpidec}. The overall
momentum distribution is changed only slightly as may be seen from the 
Table~\ref{tab:meanmom}. The values given in this table are the
averages for the interval $0 \div 30~TeV$ to cut away the extreme fluctuations.
Close values for $QGP(0.4,0.06)$ and $QGP(0.4,0.12)$ are given by the
momentum cutoff - in the case of $QGP(0.4,0.06)$ one can observe
underground only the tail of the distribution.
\newlength{\landscapeheight}
\setlength{\landscapeheight}{\textwidth}
\begin{figure}[htbp]
  \begin{center}
    \begin{sideways}
      \begin{minipage}[c]{\textheight}
        \begin{center}
          \epsfig{file=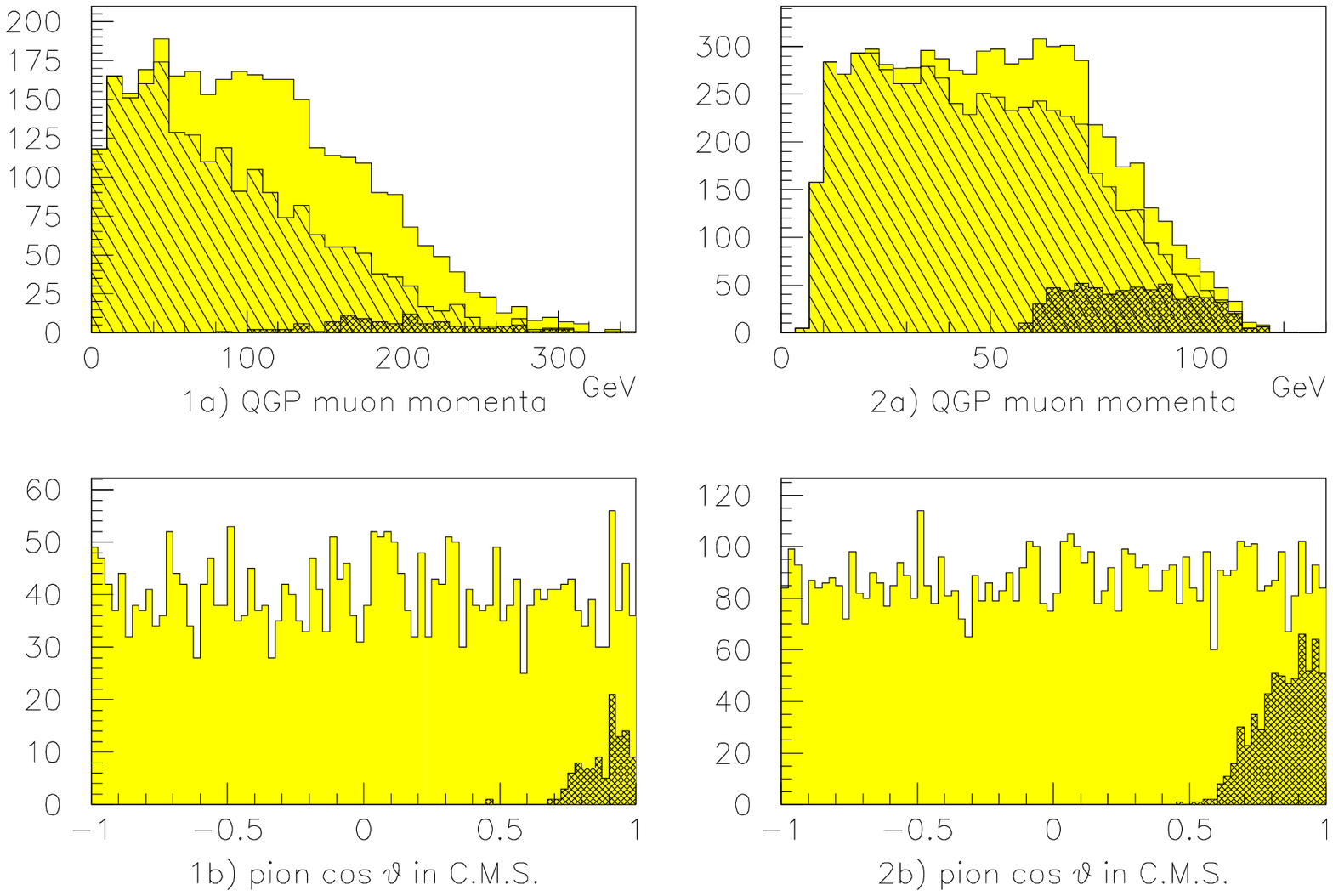,height=0.95\landscapeheight}
          \caption{Picture 1a,b): QGP(0.4,1.0); 2a,b): QGP(0.4,0.06).
            1,2a) - gray: all possible muons; hatched: muons from
            decays within 7~km from initial interaction; black:
            detected muons. 1,2b) gray: all QGP pions; black: pions
            whose daughter muons are detected.}
          \label{fig:qgmpidec}
        \end{center}
      \end{minipage}
    \end{sideways}
  \end{center}
\end{figure}
\begin{table}\centering
\begin{tabular}{|l|c|c|}   \hline
   model      &   model only  & QGP + VENUS \\ 
              &   [$GeV$]     &  [$GeV$] \\ \hline
QGSJET        & 287.5 (265.5) & - \\
VENUS         & 287.2 (259.8) & - \\
QGP(0.4,0.06) & ~99.9 ~(31.1) & 277.8 (225.8) \\
QGP(0.4,0.12) & 105.0 ~(33.6) & 277.9 (226.8) \\
QGP(0.4,0.3)  & 130.3 ~(41.4) & 279.3 (232.6) \\
QGP(0.4,1.0)  & 234.5 ~(73.7) & 282.2 (244.8) \\ \hline
\end{tabular}
\caption{Mean momentum of muons detected underground in iron initiated 
         showers. The R.M.S. values are given in the parentheses.} 
\label{tab:meanmom}
\end{table}

\begin{table}\centering
\begin{tabular}{|l|c|c|}   \hline
           &              & fraction of events \\
\multicolumn{1}{|c|}{model}&average $N^{u}_{\mu}$ & with $N^{u}_{\mu} > 990$ \\
           &                        &  (in \%) \\ \hline
VENUS~~~~~~$p$ & 464 &  ~6.3 $\pm$ 0.6 \\
QGSJET~~~~$p$ & 378 &  ~3.7 $\pm$ 0.4 \\ \hline
VENUS~~~~~$Fe$ & 807 & 20.1 $\pm$ 1.1 \\
QGSJET~~~$Fe$ & 709 & 13.6 $\pm$ 0.9 \\ \hline
QGP(0.1,0.06) & 849 & 21.8 $\pm$ 1.2 \\
QGP(0.1,0.12) & 843 & 21.7 $\pm$ 1.2 \\
QGP(0.1,0.3)  & 829 & 20.4 $\pm$ 1.1 \\
QGP(0.1,1.0)  & 819 & 20.4 $\pm$ 1.1 \\ \hline
QGP(0.2,0.06) & 868 & 23.4 $\pm$ 1.2 \\
QGP(0.2,0.12) & 871 & 22.9 $\pm$ 1.2 \\
QGP(0.2,0.3)  & 867 & 22.5 $\pm$ 1.2 \\
QGP(0.2,1.0)  & 809 & 19.3 $\pm$ 1.1 \\ \hline
QGP(0.4,0.06) & 967 & 27.0 $\pm$ 1.3 \\
QGP(0.4,0.12) & 955 & 26.5 $\pm$ 1.3 \\
QGP(0.4,0.3)  & 904 & 24.3 $\pm$ 1.2 \\
QGP(0.4,1.0)  & 835 & 22.0 $\pm$ 1.2 \\ \hline
\end{tabular}
\caption{Average underground muon size $N^{u}_{\mu}$ of events and
 the fraction of events with $N^{u}_{\mu} > 990$ . All QGP simulations
 were supplemented by VENUS.} 
\label{tab:ugsize}
\end{table}

\begin{figure}\centering
\epsfig{file=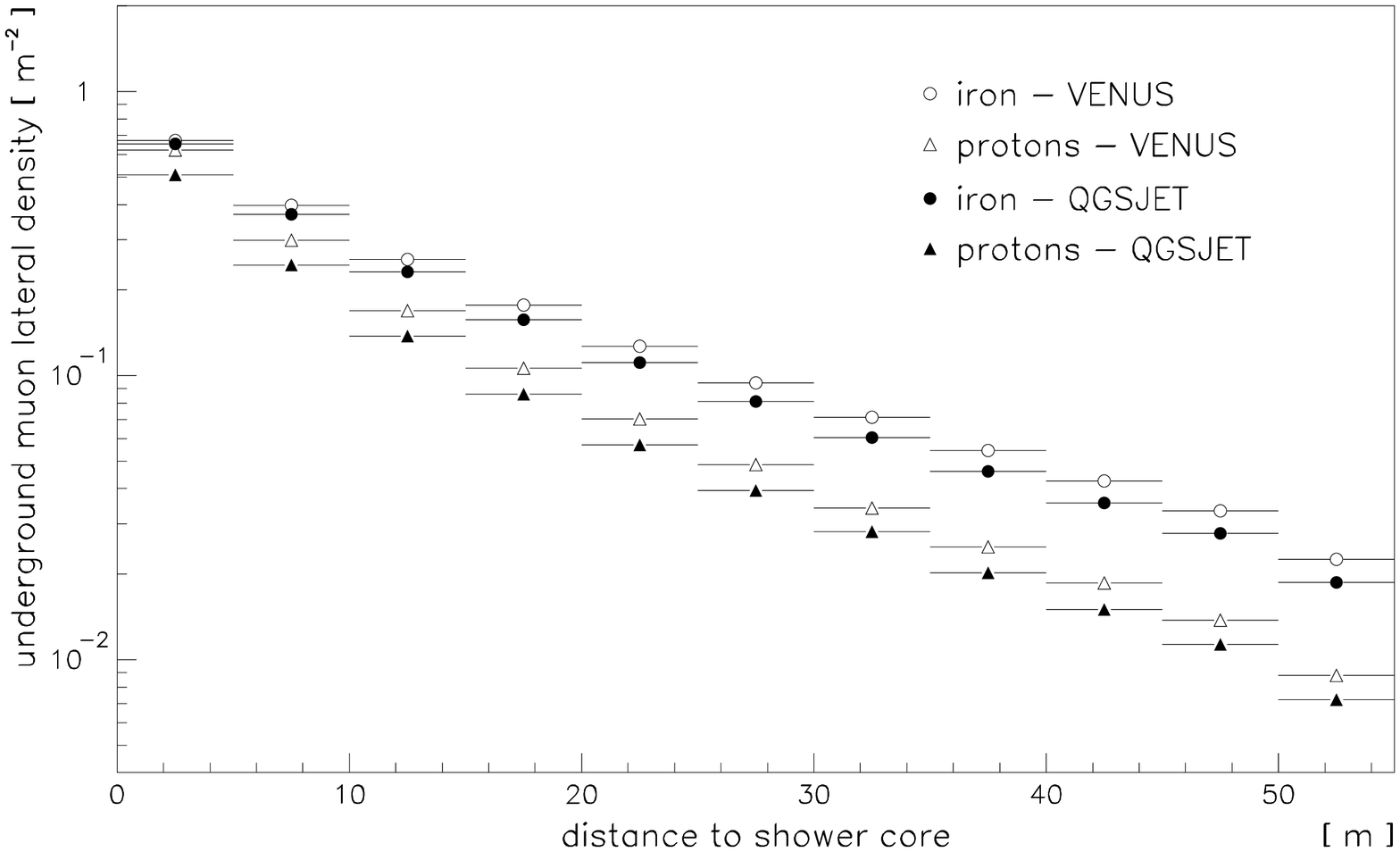,width=15cm,height=9cm}
\caption{Lateral distributions for muon densities underground produced by 
         VENUS and QGSJET.} 
\label{fig:uldfm}
\end{figure}

The underground muon size $N^{u}_{\mu}$ (necessarily truncated
by the design of experimental area) reveals notable
differences already in the case of unmodified models (Table~\ref{tab:ugsize}). 
The value used to define the tail of the underground muon size
distribution, 990 muons, has been chosen so that in the case of pure 
VENUS simulations 400 events pass the limit and the statistical error 
is only 5 \%. The effects of QGP are clearly detectable when the fraction
of melted nucleons becomes larger. The muon size at medium depth 
underground also shows differences between VENUS and
QGSJET in the case of proton showers. The muon LDF 
obtained from VENUS have systematically higher values, both for iron
nuclei and protons, than those generated by QGSJET (Fig.~\ref{fig:uldfm}). 
In all cases compared here, the statistical errors are smaller than the 
marker size. Apparently such an underground experiment can
essentially help to tune the models which is an  indispensable condition
for any further studies on chemical composition of cosmic rays and
search for QGP effects in cosmic ray showers. 

\begin{figure}\centering
\epsfig{file=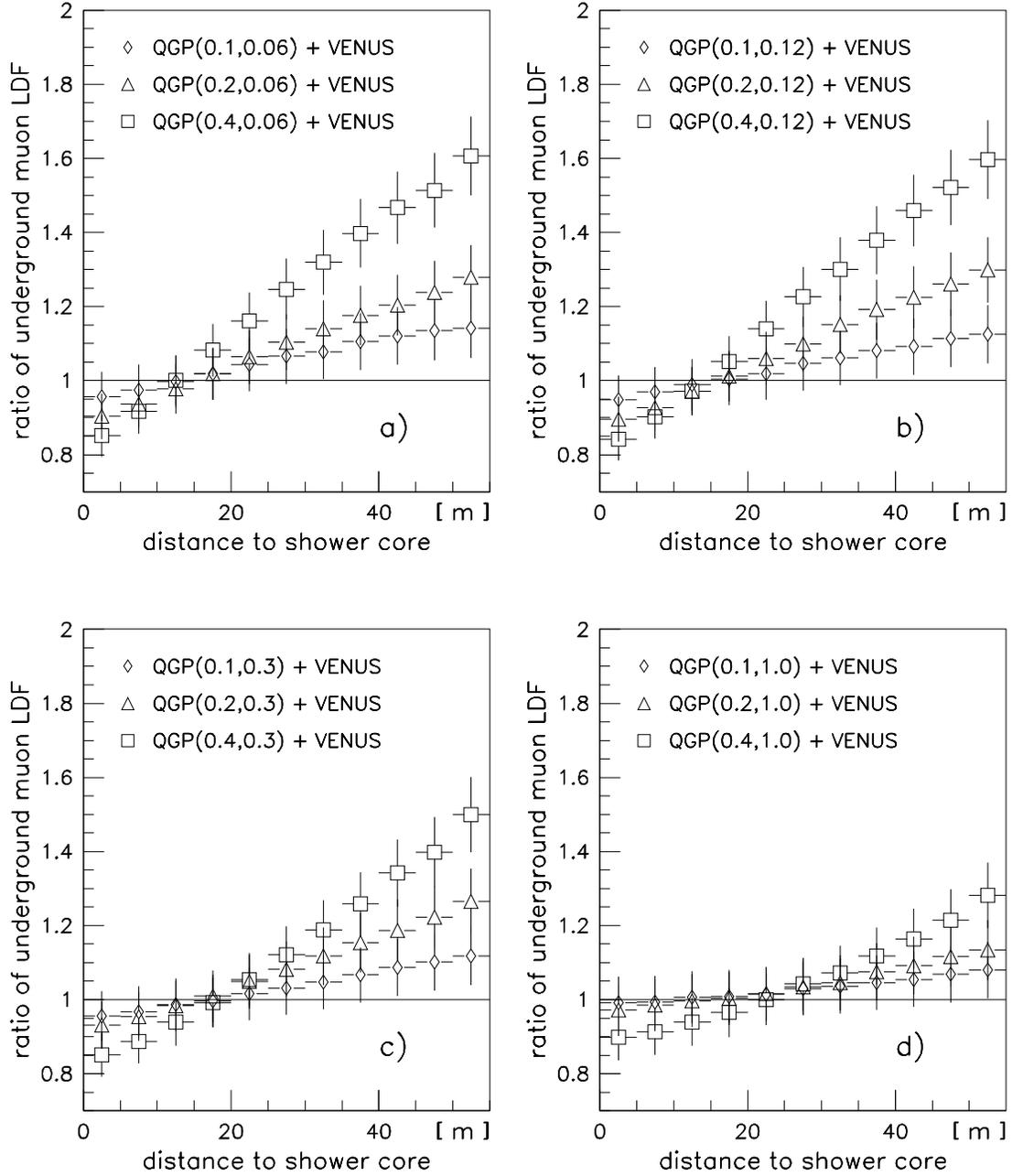,width=15cm,height=20cm}
\caption{Ratio of underground lateral muon densities with QGP simulation to
simulations done by VENUS only (events with $N^{u}_{\mu}~>~990$).} 
\label{fig:uldfrat}
\end{figure}

The underground muon LDF with QGP simulation differs from the reference 
LDF obtained from data simulated by VENUS code only. 
Fig.~\ref{fig:uldfrat} a)-d) show the underground muon lateral 
distribution functions corresponding to different values of the ratio 
${\cal R}_{h}$ divided by the reference LDF. Each picture
summarises the distributions with the same value of $<p^{\pi}\!\!>$. The
functions correspond to the events with large underground muon size -
$N^{u}_{\mu} > 990$, where the differences were most notable. The LDF
of VENUS and QGSJET with the same condition, $N^{u}_{\mu} > 990$, are
mutually much less different than the LDF without any conditions in 
Fig.~\ref{fig:uldfm} - within the statistical errors they practically 
coincide over the whole range $0 \div 55 ~m$. Hence the effects caused 
by QGP production are in this sense model independent. 
Fig.~\ref{fig:uldfrat} a)-d) show that the effects of more abundant pion
production caused by increasing ${\cal R}_{h}$ can be offset by 
increasing $<p^{\pi}\!\!>$ to a certain extent. Near the centre of 
the shower, both scenarios give similar results. But at greater distances 
from the shower core the LDF with QGP contribution decreases slower than the 
reference one and starts to differ. This 'point of departure' moves away
from the shower centre with increasing $<p^{\pi}\!\!>$ and becomes more
prominent with the increase of ${\cal R}_{h}$.

As we are not dealing with real data it is not considered necessary
to increase 
the statistics to pinpoint exactly the values of ${\cal R}_{h}$ and 
$<p^{\pi}\!\!>$ at which the distinction between models with and without 
QGP can be made. The comparison of the LDF's shows that with increasing
$<p^{\pi}\!\!>$ one has also to increase the distance upto which the LDF can
be measured in order to be able to see possible QGP effects. At small 
${\cal R}_{h}$ one has to increase the observation time or the detection 
area to achieve the same result. Thus the resolving power of the experiment 
is related to the dimensions of the detector available underground.
Assuming the detection area to be $100 \times 100 \; m^{2}$ and 10 \% 
iron nuclei and 90 \% protons in the primary flux, one can get in 
$2.4~10^{7}~sec$ (75 \% of a year) $15~000$ iron initiated events and 
34 000 proton initiated events, with $N^{u}_{\mu} > 990$.
These numbers correspond to showers initiated by
particles with $E_{inc} > 10^{15}~eV$. In the case of $E_{inc} > 10^{16}~eV$ 
the overall flux is smaller and one would get only about 200 iron events
and 400 proton events, assuming the same fraction for events above 990 muons
cut as at $E_{inc} > 10^{15}~eV$. Actually the percentage of events above 
this cut increases with energy and the yield of observed events
should be higher.

\begin{figure}\centering
\epsfig{file=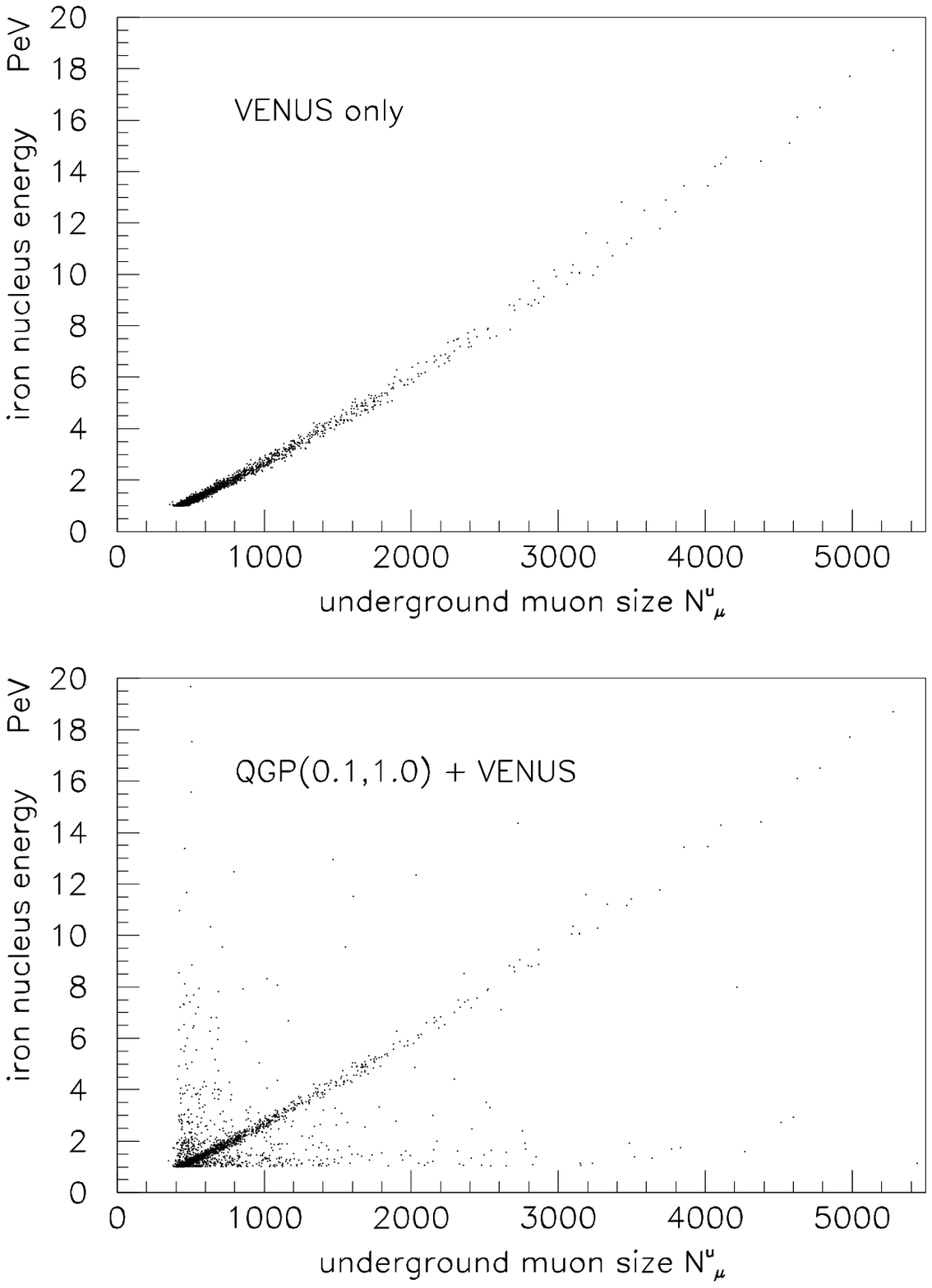,width=15cm,height=20cm}
\caption{Underground muon size versus primary energy of iron nuclei.} 
\label{fig:edefic}
\end{figure}

\section{Discussion} %
\label{secD}

    Results obtained from simulations discussed above cannot be compared at 
present with real data as the muon detectors are either located at ground 
level where the high energy component is entirely hidden in the overall 
muon flux or they are located so deep underground that the muon
momentum cutoff is of the order of TeV or larger and only few muons
penetrate to the detector (\cite{MACRO},\cite{SOUDAN},\cite{SCE},
\cite{NUSEX},\cite{LVD},\cite{Frejus}). The few experiments with medium 
overburden have either crude spatial resolution like the underwater 
experiments \cite{Bugaev} and \cite{BAKSAN} or  they have relatively 
small detection area like \cite{KGF} ($6 \times 6 ~m^{2}$).

The present study is based on the assumption that the high energy 
interactions of cosmic rays in the atmosphere may lead to the
production of QGP in the initial stage of the shower development.
The inclusion of a very simple model of QGP production and freeze-out into
high energy interactions of iron nuclei with air in cosmic ray showers has
shown that under certain kinematical conditions one could observe the QGP
induced effects in the high energy muon component of extensive air showers.
These kinematical conditions are mainly the availability of sufficient 
energy $E_{hot}$ of melted nucleons and not too high mean momentum 
$<p^{\pi}\!\!>$ of the evaporated pions.  The model has been intentionally 
chosen so simple so that it can cover sufficiently large interval of these 
two quantities with minimum parameters. 
Moreover the simulations have shown that the underground data on muons are
rather insensitive to the isotropy of initial QGP pion distribution 
(Fig.~\ref{fig:qgmpidec}).
It is interesting that none of the simulated QGP systems increases the 
muon density in the very centre of the shower and this number is 
essentially determined by the unmodified high
energy interaction model (VENUS in this case).

Even in the situation when there is no apparent difference in the
underground muon size $N^{u}_{\mu}$ or underground LDF, like in the 
case of $QGP(0.1,1.0)$, the QGP production can still be detected
provided there is an independent measurement of the shower energy, 
e.g. the muon size $N^{tr}_{\mu}$ on the surface. This quantity is 
proportional to the energy of the incident nucleus and the 
proportionality holds even underground (Fig.~\ref{fig:edefic}). Since 
the creation of QGP distorts the $energy-muon~size$ relation, a
comparison of the muon sizes on the surface and underground may 
indicate deviation from the standard scenario. Events under the 
diagonal of proportionality $E_{inc} \sim N^{u}_{\mu}$
are what one would expect - as the abundant pion production in QGP 
mimics higher energy of the projectile nucleus. 
The events above the diagonal are more intriguing. 
In this case clearly the muons produced from QGP are not able to
offset the loss which is due to several factors. One
of them is that in normal interactions the leading particle effect
constrains the lateral energy dissipation. No such effect has been
included into QGP simulation and this combined with sometimes very high
altitude of the primary interaction leads to the spraying of the 
QGP originated muons on a very large area. Moreover the pions from QGP 
mostly do not start particle cascades and their energy is not converted into 
secondary particles any more. Thus one can take roughly the muon 
multiplicity deficit in a standard interaction as 
$ N_{deff} = k_{std} \cdot E_{hot} - q_{std}$ while
the gain in multiplicity from QGP is 
$N_{QGP} = (k_{QGP} \cdot E_{hot} - q_{QGP}) \cdot f_{det}$, 
where $f_{det}$ is the detection efficiency factor which is quite 
small (see Fig.~\ref{fig:qgmpidec}),
$k_{QGP}$ and $k_{std}$ are the slope parameters and $q_{QGP}$ and 
$q_{std}$ are the intercepts.
The net result $N_{QGP} - N_{deff}$ is apparently a linear function with
positive intercept and negative slope. At smaller $E_{hot}$ while
the intercept prevails one gets higher multiplicity than normal and when
higher energy portion is lost from standard production one gets events
with lower muon multiplicities underground. 

Another factor
contributing to the multiplicity deficit is the exemption of $N_{hot}$
nucleons from further interactions - there are only $56 - N_{hot}$
nucleons with the energy $E_{inc} \cdot (56-N_{hot})/56$. The baryons
evaporated from QGP cannot provide enough muons to balance the loss of
forward production initiated by $N_{hot}$ nucleons in standard
interaction. This part of the deficit is quite difficult to express
quantitatively in terms of linear relations. Otherwise it
would be much easier to measure chemical composition of cosmic rays !
In fact, events with lower muon multiplicity underground
could be mistaken as products of interactions of lighter primary nuclei.
Therefore, the resolution in energy measurement both at surface and 
underground will be crucial to make the distinction.

    The measurements on energetic muons in cosmic ray showers at medium 
depth underground can reveal information on the initial stages 
of interactions which are otherwise inaccessible through measurements on 
the surface. Provided the QGP is formed in interactions of lighter nuclei, 
the number of produced muons will grow linearly with c.m.s. energy per 
nucleon. The ratio ${\cal R}_{h}$ should, in naive expectation, also grow with
the energy. Hence the observable effects are likely to become more apparent 
at higher energies (in the simple model used here, ${\cal R}_{h}$ was 
considered to be energy independent). Thus, in principle, one should be 
able to study some of the scenarios of QGP deflagration or detonation at 
the freeze out described in \cite{PRC}. 

\vspace{1cm}I would like to thank Rupert Leitner and Suresh Tonwar
for valuable comments. I am very grateful to Jiri Masik for his continuing
and patient help in solving all problems connected with computing. The work 
was supported by the Ministry of Education of the Czech Republic within
the project LN00A006 and by the grant A1010928/1999 of GA AV CR.



\end{document}